\documentclass{article}

\usepackage{PRIMEarxiv}

\usepackage[utf8]{inputenc} % allow utf-8 input
\usepackage[T1]{fontenc}    % use 8-bit T1 fonts
\usepackage{hyperref}       % hyperlinks
\usepackage{url}            % simple URL typesetting
\usepackage{booktabs}       % professional-quality tables
\usepackage{amsfonts}       % blackboard math symbols
\usepackage{nicefrac}       % compact symbols for 1/2, etc.
\usepackage{microtype}      % microtypography
\usepackage{lipsum}
\usepackage{fancyhdr}       % header
\usepackage{graphicx}  
\usepackage{multirow}
\usepackage{multicol}
\usepackage{amsmath,amssymb,amsfonts}
\usepackage{amsthm}
\usepackage{mathrsfs}
\usepackage{xcolor}
\usepackage{textcomp}
\usepackage{manyfoot}
\usepackage{booktabs}
\usepackage{algorithm}
\usepackage{algorithmicx}
\usepackage{algpseudocode}
\usepackage{listings}
\usepackage{todonotes}
\usepackage{changes}
\usepackage[font=scriptsize]{subcaption}
\graphicspath{{media/}}     % organize your images and other figures under media/ folder

\newcommand{\vx}{{\mathbf{x}}}
\newcommand{\vy}{{\mathbf{y}}}

\newcommand{\on}{{\text{on}}}

\newcommand{\eref}[1]{Eq.~\eqref{eq:#1}}
\newcommand{\fref}[1]{Fig.~\ref{fig:#1}}

\newcommand{\aref}[1]{\algo.~\ref{algo:#1}}

\newcommand{\algo}{\textbf{Algorithm}}

\def\figclip#1#2#3{\includegraphics[trim=#1, clip, width=#2\columnwidth]{#3}}

\title{Online t-SNE for single-cell RNA-seq}

\author{
  Hui Ma, Kai Chen* \\
 }
\begin{document}
\maketitle

\begin{abstract}
Due to the sequential sample arrival, changing experiment conditions, and evolution of knowledge, the demand to continually visualize evolving structures of sequential and diverse single-cell RNA-sequencing (scRNA-seq) data becomes indispensable. However, as one of the state-of-the-art visualization and analysis methods for scRNA-seq, t-distributed stochastic neighbor embedding (t-SNE) merely visualizes static scRNA-seq data offline and fails to meet the demand well. To address these challenges, we introduce online t-SNE to seamlessly integrate sequential scRNA-seq data. Online t-SNE achieves this by leveraging the embedding space of old samples, exploring the embedding space of new samples, and aligning the two embedding spaces on the fly. Consequently, online t-SNE dramatically enables the continual discovery of new structures and high-quality visualization of new scRNA-seq data without retraining from scratch.
We showcase the formidable visualization capabilities of online t-SNE across diverse sequential scRNA-seq datasets. 
\end{abstract}

\keywords{Single-cell RNA-sequencing \and t-SNE \and continual
visualization \and embedding space }

\section{Introduction}
{S}ingle-cell RNA-sequencing (scRNA-seq) has revolutionized the study of gene expression \cite{poran2017single, travaglini2020molecular, xue2013genetic, mah2024cell, grabski2023significance} by enabling the profiling of individual cells within heterogeneous biological samples, offering a profound glimpse into the intricacies of cell biology. The high-dimensional space of scRNA-seq encapsulates the transcriptional activity of thousands of genes, making efficient analysis and visualization paramount. The visualization and analysis of scRNA-seq data have ushered in a new era of biological discovery \cite{li2022universal, marouf2020realistic, zhao2023single, narayan2021assessing, ozgode2023effect, wang2023pairing}. 
Integrative visualization and analysis of diverse high-dimensional scRNA-seq enable the understanding and discovery of cell structures and constituents \cite{costa2021demonstrating,perez2020improving}. 
In this context, various visualization and dimensionality reduction methods, including principal component analysis(PCA) \cite{zuo2006bidirectional,granato2018use}, multidimensional scaling(MDS) \cite{vesanto2000clustering, torgerson1952multidimensional}, and self-organizing map(SOM) \cite{kohonen1990self, kohonen2013essentials}, have been employed for single-cell genomics \cite{kobak2019art,yao2012independent,jakaitiene2016multidimensional,tamayo1999interpreting}. 
In addition, popular emerging dimensionality reduction methods for single-cell genomics include locally linear embedding(LLE) \cite{roweis2000nonlinear, li2008locally}, Laplacian eigenmaps \cite{belkin2003laplacian, belkin2003laplacian},  maximum entropy unfolding, uniform manifold approximation and projection (UMAP)\cite{mcinnes2018umap, pealat2021improved}, and t-distributed stochastic neighbor embedding (t-SNE) \cite{van2008visualizing, chatzimparmpas2020t}. 
These approaches are not solely tasked with the elucidation of cellular composition and diversity within a sample but also facilitate the identification of cell clusters and the exploration of expression dynamics. 

Compared with them, t-SNE \cite{van2008visualizing} undoubtedly stands out as one of the most effective nonlinear visualization techniques of high-dimension scRNA-seq data.  t-SNE distills the complexity of scRNA-seq heterogeneity into understandable representations.  By embedding the profiling of individual cells within heterogeneous biological samples, transformative t-SNE offers a powerful lens through which to explore the intricacies of cell-type diversity, dynamic cell states, and the underlying molecular mechanisms governing a wide array of biological processes. 
The core concept of t-SNE is to convert the similarity between high-dimension samples into affinities 
and preserve the similarity between low-dimension points in the embedding space as much as possible. 
t-SNE computes the affinity 
by using a nonlinear kernel function. t-SNE can effectively mitigate the effects of dimensionality catastrophe and render it adaptable for scRNA-seq data with complex structures. While there are many extensions of t-SNE.
A substantial portion of t-SNEs has primarily focused on parameter tuning \cite{kobak2021initialization}, computation efficiency \cite{van2014accelerating,linderman2019fast,meyer2020improving}, and multiple applications \cite{kang2021conditional,pezzotti2016hierarchical,de2018perplexity}.

Due to the consecutive generation of scRNA-req data, visualization and analysis of sequential scRNA-seq data becoming essential in the field of single-cell genomics, including tasks such as cell classification, sub-population identification, and cell trajectory analysis. 
In analyzing sequential scRNA-seq data, the focus shifts from static snapshots of cellular states to the dynamic transitions and trajectories that cells traverse over time. 
At its core, to extract meaningful insights from such sequential scRNA-seq data, a robust online method has become imperative, which can decipher cell composition and diversity. 
To capture sequential dependencies in sequential scRNA-seq data, online visualization and analysis of sequential scRNA-req data presents several unique challenges that distinguish it from existing static variants of t-SNE. 
Given the embedding of old scRNA-req data, existing variants of t-SNE face the challenge of exploring the embeddings of new data due to a lack of sequential dependencies. In other words, all existing variants of t-SNE run in an offline manner. The rapid advancement of single-cell transcriptomics demands the development of corresponding computational online visualization and analysis algorithms to discover and extract integrative insights from extensive and diversified sequential scRNA-seq data. This discrepancy underscores the necessity for online t-SNE that overcomes the limitations of popular offline t-SNE.
We observe some of the key challenges for t-SNE as follows:

\begin{itemize}
\item \textbf{Concept drift:} secRNA-seq data is often generated in dynamic experiment conditions where the underlying data distribution can vary over time. 
This situation, known as concept drift, leads to decreasing visualization accuracy as time progresses. 

\item \textbf{Efficient computational visualization:} As sequential secRNA-seq data arrives continuously and the number of samples grows with time, efficient visualization is crucial and troublesome for current offline t-SNE. 

\item \textbf{Discovering evolving structures:} 
When we receive scRNA-req data sequentially, static and offline t-SNE cannot effectively capture sequential dependencies between the subsets of old data and new data, and explore the embeddings of new data.
\end{itemize}

We delve into the core principles, challenges, and profound implications of online t-SNE.
In contrast to traditional t-SNEs, online t-SNE blurs the boundaries between the learning and visualization phases. 
Online t-SNE, designed to process data one subset by one subset, offers significant gains in efficiency, both in terms of time and space utilization, making them an exceptionally practical alternative to visualization and analysis of sequential scRNA-seq data. 
Notably, it exerts a growing influence on the theoretical and algorithmic advancements within the field of single-cell genomics. 
Specifically, our main contributions are as follows:
\begin{enumerate}
    \item[1)] Online t-SNE can adapt to evolving data distributions or patterns through the inclusion of new samples, enabling seamless online integration of sequential scRNA-seq data.
 
    \item[2)] We provide a unified and compositional high-dimensional joint probability using Gaussian distribution, to incorporate the similarity between old samples, connect old and new samples, and describe the similarity between new samples of sequential scRNA-seq data.
        
    \item[3)] We introduce a unified and compositional low-dimensional joint probability using Student-$t$ distribution, to integrate the embeddings of old samples, transfer knowledge between embeddings of old and new samples, and encode the embeddings of new samples.        
        
    \item[4)] We design an online Kullback-Leibler (KL) divergence to minimize the inconsistency between the compositional high-dimensional joint probability and the compositional low-dimensional joint probability, and capture the sequential dependency between old and new scRNA-seq data. 

    \item[5)] Online t-SNE explores shared embedding across diversified subsets of new data, discovers new biological structures from new data, and provides immediate insights and visual understanding of the data.
\end{enumerate}

Overall, online t-SNE facilitates dynamic and continual visualization, enabling us to capture the temporal evolution of data clusters and relationships.

\section{Theory}
\subsection{t-distributed stochastic neighbor embedding}
We first briefly introduce the principle of standard t-SNE \cite{van2008visualizing,cai2022theoretical}, its basic notation, definition, and algorithm pipeline. In the following sections, we will use standard t-SNE and offline t-SNE interchangeably due to their equivalence in visualizing sequential scRNA-seq data. 
Offline t-SNE begins by transforming the high-dimensional Euclidean distances among samples into conditional probabilities, reflecting their similarities. 
Offline t-SNE evolved from its predecessor, a method known as stochastic neighbor embedding (SNE). The core concept of SNE involved characterizing the relationships between pairs of high-dimensional points through normalized affinities. Close neighbors exhibit high affinity, while distant samples have near-zero affinity. SNE then arranged the points in a two-dimensional space, aiming to minimize the KL divergence between the affinities in the spaces of high and low dimensions.

\begin{algorithm}[!h]
\renewcommand{\algorithmicrequire}{\textbf{Input:}}
\renewcommand{\algorithmicensure}{\textbf{Output:}}
\caption{Offline t-SNE} \label{algo:tsne}
\begin{algorithmic}[1]
\Require  high dimensional samples $X=[\vx_{1}, \vx_{2}, \dots, \vx_{n}]^{\top}$, hyperparameters including perplexity $\mathrm{Perp}$, number of iterations $T$, learning rate $\eta$, and momentum $\alpha(t)$. 
    \Ensure  low dimensional embedding $Y^{(T)}=[\vy_{1}, \vy_{2}, \dots, \vy_{n}]^{\top}$.
    \State Compute pairwise affinities $p_{j|i}$ with perplexity $\mathrm{Perp}$;
    \State Set $p_{ij}=\frac{p_{i|j}+p_{j|i}}{2n}$;
    \State Initialize $Y^{(0)}=[\vy_{1}, \vy_{2}, \dots ,\vy_{n}]^{\top}$ using PCA;
    \For{$t=1$ to $T$}
    \State Compute low-dimensional affinities $q_{ij}$;
    \State Compute gradient $\frac{\partial C}{\partial \vy}$;
    \State Set ${{\vy}^{(t)}}={{\vy}^{(t-1)}}+\eta \frac{\partial C}{\partial \vy}+\alpha (t)({{\vy}^{(t-1)}}-{{\vy}^{(t-2)}})$.
    \EndFor
    \end{algorithmic}
\end{algorithm}

Specifically, given a high-dimensional dataset $X=[{{\vx}_{1}},{{\vx}_{2}},...,{{\vx}_{n}}]^{\top}$, where $\vx\in{\mathbb{R}}^{D}$ and $n$ is the number of samples, t-SNE aims to find the corresponding low-dimensional embedding $Y=[{{\vy}_{1}},{{\vy}_{2}},...,{{\vy}_{n}}]^{\top}$ for $X$, where $\vy\in {\mathbb{R}^{d}}$ and $d<D$. Within t-SNE, if samples $\vx_{i}$ and $\vx_{j}$ are close in the high-dimensional space, their low-dimensional embeddings $\vy_{i}$ and $\vy_{j}$ should be close. 
The likeness between samples $\vx_{i}$ and $\vx_{j}$ is expressed through the conditional probability, $p_{j|i}$. This probability signifies the likelihood that $\vx_{i}$ would choose $\vx_{j}$ as its neighbor if neighbors were selected based on their probability density under a Gaussian centered at $\vx_{i}$. In the case of closely situated samples, $p_{j|i}$ tends to be relatively high, while for significantly distant samples, $p_{j|i}$ becomes nearly infinitesimal.
The pairwise similarity (denoted by $p_{ij}$ ) between samples $\vx_{i}$ and $\vx_{j}$ in the high-dimensional space is defined as
\begin{align} \label{eq:tpij} 
p_{ij} &= \frac{p_{j|i} + p_{i|j}}{2n},
\end{align}
where
\begin{align} \label{eq:pji}
p_{j|i} &= \frac{\exp \big(-\frac{\|\vx_{i} - \vx_{j}\|^2}{2\sigma_i^2}\big)}{\sum\limits_{k\ne i} \exp \big(-\frac{\|\vx_{i} - \vx_k\|^2}{2\sigma_i^2}\big)}.
\end{align}
A crucial parameter involved in \eref{pji} is the variance $\sigma_{i}$ of the Gaussian. 
Since data density varies across the dataset, it is important to choose a reasonable value for  $\sigma_{i}$.
The variance $\sigma_{i}$ is typically selected by presetting the perplexity of the distribution.
The low-dimensional counterparts, $\vy_i$ and $\vy_j$, corresponding to the high-dimensional samples $\vx_i$ and $\vx_j$, allow for the computation of a similar conditional probability denoted as $q_{j|i}$. Thus, we characterize the similarity between $\vy_i$ and $\vy_j$ by using the $t$-distribution with one degree of freedom:
\begin{align} \label{eq:qji}
	q_{ij} &= \frac{\big(1+{\|\vy_{i} - \vy_{j}\|^2}\big)^{-1}}{\sum\limits_{k\ne l}\big(1+{\|\vy_{k} - \vy_l\|^2}\big)^{-1}}.
\end{align}
If the low dimensional points $\vy_i$ and $\vy_j$ accurately represent the similarity between the high-dimensional samples $\vx_i$ and $\vx_j$, the conditional probabilities $p_{ij}$ and $q_{ij}$ will be identical. Inspired by this insight, t-SNE discovers a low-dimensional data embedding that minimizes the discrepancy between $p_{ij}$ and $q_{ij}$. A natural measure of how faithfully $q_{ij}$ models $p_{ij}$ is the KL divergence across all samples. The cost of offline t-SNE for minimizing the KL divergences is defined as follows:
\begin{align}\label{eq:t cost}
\begin{split}
    C_{\text{off}}=\sum\limits_{i}\mathrm{KL}({{P}_{i}}\|{{Q}_{i}})=\sum\limits_{i, j}{p}_{ij}\log \frac{{p}_{ij}}{{q}_{ij}}.
\end{split}
\end{align}
We can obtain the minimizing by using a gradient descent approach. This minimizing significantly punishes employing widely separated $\{\vy_{i}, \vy_{j}\}$ to represent nearby samples $\{\vx_{i}, \vx_{j}\}$.
the learning of offline t-SNE is described in \aref{tsne}.
Note that the computation complexity of the t-SNE algorithm is $O(n^{2})$.

\subsection{Online t-distributed stochastic neighbor embedding} 

While offline t-SNE produces reasonably effective visualizations, it grapples with a challenge to continually learn the embedding of sequential data and a predicament known as the "online problem with dynamical data". 
The online t-SNE deviates from the offline SNE in three key ways: (1) it adopts an online conditional probability for high dimensional samples, as briefly indicated by compositional Gaussian distribution, (2) it employs an online conditional probability for low dimensional embeddings, as denoted by compositional Student-$t$ distribution, and (3) it enjoys an online KL divergence to gauge the discrepancy between the aforementioned two probabilities.
Furthermore, we first treat the sequential data as a set of old and new data, where the old data subset consists of historical samples and the new data subset denotes the newly collected or incoming samples. The standard t-SNE is offline because it merely learns the low-dimensional embedding of old data.
	
\subsubsection{Compositional high-dimensional probabilities using Gaussian distribution}
For existing methods of t-SNE\cite{lee2015multi,linderman2019clustering}, 
once the pairwise similarity is computed, it will be fixed and frozen throughout the optimization to obtain the representation of corresponding low-dimensional embedding. 
However, the characteristic of pairwise similarity in offline t-SNE cannot describe the similarity between the samples of old data and new data. In particular, this makes t-SNE fail to adapt to the evolution of sequential scRNA-seq data always including incoming new samples.
For online t-SNE, we follow the form of the pairwise similarity ${p}_{ij}$ in offline t-SNE. To accurately reflect the pairwise similarity across old samples and new samples, we propose a compositional high-dimensional joint probability to describe the similarity for all sequential scRNA-seq data.
Specifically, the compositional high-dimensional joint probability includes three types of components: (1) a joint probability ${p}_{ij}$ on old samples, (2) a joint probability ${p}_{ij_{*}}$ connecting old and new (with subscript $*$) samples, and (3) a joint probability ${p}_{i_{*}j_{*}}$ on new samples. Therefore, there are three types of high dimensional similarities (denoted by $\tau^{H}$) based on Gaussian distribution: (1) old similarity $\tau^{H}_{ij}=\exp(-\|{\vx_{i}}-{\vx_{{{j}}}}\|^{2}/2\sigma_{i}^{2})$ between old samples $\vx_i$ and $\vx_j$, (2) cross similarity $\tau^{H}_{ij_{*}}=\exp(-\|{\vx_{i}}-{\vx_{{{j}_{*}}}}\|^{2}/2\sigma_{i}^{2})$  between old sample $\vx_i$ and new sample  $\vx_{j_{*}}$, where subscript $j_{*}$ is the index of 
$\vx_{j_{*}}$  (3) new similarity $\tau^{H}_{i_{*}j_{*}}=\exp(-\|\vx_{{i}_{*}}-\vx_{{j}_{*}}\|^{2}/2\sigma_{{i}}^{2})$ between new samples $\vx_{i_{*}}$ and $\vx_{j_{*}}$.
The first component, namely, joint probability on old samples, is defined similarly with standard t-SNE (see \eref{tpij}). For the second component, we define the joint probability connecting old and new samples as follows:
\begin{align}\label{eq:HGD-old}
    {{p}_{ij_{*}}} &=\frac{{{p}_{i|j_{*}}}+{{p}_{j_{*}|i}}}{n+m},
\end{align}
where $n$ and $m$ denote the numbers of old samples and new samples, respectively.
Furthermore, we define the conditional probabilities 
${{p}_{{{j}_{*}}|i}} $ as follows:
\begin{align}\label{eq:HGD-cross}
{{p}_{{{j}_{*}}|i}} &=
\frac{\tau^{H}_{ij_{*}}}{\sum\limits_{k\ne i}\tau^{H}_{ki}+\sum\limits_{k_{*}}\tau^{H}_{k_{*}i}}, \\
{p}_{j|i_{*}} &=
\frac{\tau^{H}_{ij_{*}}}{\sum_{k_{*}\ne i_{*}}\tau^{H}_{k_{*}i_{*}} + \sum\limits_{k}\tau^{H}_{ki_{*}}}.
\end{align}
This ${{p}_{{{j}_{*}}|i}}$ transforms the high-dimensional Euclidean distances between old and new samples into conditional probabilities that depict their temporal relationship. 
${{p}_{{{j}_{*}}|i}}$ represents the likelihood that $\vx_{j_{*}}$ would select $\vx_{i}$ as its neighbor if neighbors were chosen based on their probability density under a Gaussian centered at $\vx_{i}$. $p_{{j}_{*}|i}$ plays the role of transferring knowledge between old data and new data, which characterizes the directional affinity of old sample $\vx_{j}$ to new sample $\vx_{{i}_{*}}$. 
In addition, the joint probability of new samples is defined as follows: 
\begin{align}\label{eq:HGD-new}
    {{p}_{i_{*}j_{*}}} &=\frac{{{p}_{i_{*}|j_{*}}}+{{p}_{j_{*}|i_{*}}}}{n+m},
\end{align}
where ${{p}_{j_{*}|i_{*}}}=\frac{\tau^{H}_{j_{*}i_{*}}}{\sum_{k_{*}\ne i_{*}}\tau^{H}_{k_{*}i_{*}} + \sum\limits_{k}\tau^{H}_{ki_{*}}}$.
The form of this probability encodes the underlying structure in new data. Therefore, ${p}_{ij}$ and ${p}_{i_{*}j_{*}}$ encode the high dimensional underlying structure of old data and new data, respectively. 
As new data is arriving, the computation of ${p}_{i_{*}j_{*}}$ is online and does not depend on old data. On the other hand, ${{p}_{{{j}_{*}}|i}}$ is computed online to act as a knowledge transfer bridge between old and new data.

\subsubsection{Compositional low-dimensional probabilities using Student-$t$ distribution}

In current t-SNE methods, the low-dimensional pairwise similarity fails to represent the relationship between the embeddings of historical and new samples in various sequential scRNA-seq dataset consistently introducing new samples. 
For online t-SNE, we maintain the low-dimensional pairwise similarity $q_{ij}$ on the embedding of old data from standard t-SNE. However, to accurately depict the relationship across embeddings of historical and new samples, we introduce a compositional low-dimensional joint probability to describe the similarities for all embeddings. Correspondingly, this compositional joint probability comprises three types of components: (1) a joint probability ${q}_{ij}$ for the embeddings of old data, (2) a joint probability ${q}_{ij_{*}}$ bridging the embeddings of old and new (denoted with subscript $*$) data, and (3) a joint probability ${q}_{i_{*}j_{*}}$ for the embeddings of new data. 
As a result, we have three types of low dimensional similarities (denoted by $\tau^{L}$) based on Student-$t$ distribution: (1) old similarity $\tau^{L}_{ij}={{(1+\|{\vy_{i}}-{\vy_{j}}\|^{2})}^{-1}}$ between embeddings $\vy_i$ and $\vy_j$, (2) cross similarity $\tau^{L}_{ij_{*}}={{(1+\|{\vy_{i}}-{\vy_{j_{*}}}\|^{2})}^{-1}}$ between embeddings $\vy_i$ and $\vy_{j_{*}}$, where $\vy_{j_{*}}$ is the low dimensional embedding of $\vx_{j_{*}}$, (3) new similarity $\tau^{L}_{i_{*}j_{*}}={{(1+\|{\vy_{i_{*}}}-{\vy_{j_{*}}}\|^{2})}^{-1}}$ between embeddings $\vy_{i_{*}}$ and $\vy_{j_{*}}$.
Concretely, the first component of the compositional low-dimensional probability is the same as the standard t-SNE (see \eref{qji}).
More importantly, we define the second component of the compositional low-dimensional probability as follows:
\begin{align}\label{eq:ltD-cross}
\begin{split}
    {q}_{ij_{*}} &=\frac{\tau^{L}_{ij_{*}}}{\sum\limits_{k\ne l}\tau^{L}_{kl}+\sum\tau^{L}_{kl_{*}}+
    \sum\limits_{k_{*}\ne l_{*}}\tau^{L}_{k_{*}l_{*}}}.
\end{split}
\end{align}
Furthermore, for new samples of sequential data, we can obtain the joint probability between them as follows:
\begin{align}\label{eq:ltD-new}
\begin{split}
    {q}_{{{i}_{*}}{{j}_{*}}}=\frac{\tau^{L}_{i_{*}j_{*}}}{\sum\limits_{k\ne l}\tau^{L}_{kl}+\sum\tau^{L}_{kl_{*}}+
    \sum\limits_{k_{*}\ne l_{*}}\tau^{L}_{k_{*}l_{*}}}.
\end{split}
\end{align}
These probabilities do not damage the embeddings of old data, can transfer the knowledge of the embeddings of old data to the embeddings of new data, and allow new embeddings of new samples to be discovered. The computation of new embeddings is online when new samples are arriving. 

\subsubsection{Online Kullback-Leibler divergence aligning high and low dimensional probabilities}
Here, we introduce an online KL divergence to measure the consistency between the probabilities of high-dimensional data space and low-dimensional embedding space. 
The online KL divergence serves as a quantification of the dissimilarity between three probability pairs including $\{p_{ij}, q_{ij}\}$ on old data, $\{p_{ij_{*}}, q_{ij_{*}}\}$ across old and new data, and $\{p_{i_{*}j_{*}}, q_{i_{*}j_{*}}\}$ on new data. We use $C_{1}=\mathrm{KL}(p_{ij}\|q_{ij})$,  $C_{2}=\mathrm{KL}(p_{ij_{*}}\|q_{ij_{*}})$,  and $C_{3}=\mathrm{KL}(p_{i_{*}j_{*}}\|q_{i_{*}j_{*}})$ to denote the costs of the dissimilarities, 
respectively. Note that $C_{1}$ has been minimized during the embedding learning of old data. Therefore, the online KL divergence mainly focuses on minimizing both $C_{2}$ and $C_{3}$. 
To attain a low-dimensional embedding of the new data while preserving the learned low-dimensional embedding of old data, we define the online KL divergence $C_{\on}$ as follows:
\begin{align}\label{eq:onlineKL}
\begin{split}
C_{\on}(\vy) = \overbrace{\sum_{i,j} {p_{ij}\log\frac{p_{ij}}{q_{ij}}}}^{C_{1}}
+ \overbrace{\sum_{i,{j_*}} {p_{i{j_*}}\log\frac{p_{i{j_*}}}{q_{i{j_*}}}}}^{C_{2}} 
+ \overbrace{\sum_{{i_*}{j_*}}p_{{i_*}{j_*}}\log\frac{p_{{i_*}{j_*}}}{q_{{i_*}{j_*}}}}^{C_{3}},
\end{split}
\end{align}
where $C_{1}$ is fixed.
The online KL divergence, $C_{\on}=C_{1}+C_{2}+C_{3}$,  incorporates the knowledge and embeddings of old data, connects new data and old data, and facilitates the adaption of new data. 
Using the online KL divergence, we can continually compute the low-dimensional embeddings of new samples to obtain their visualization. 
Specifically, we have the gradient of low dimensional embeddings computed by 
\begin{align}
    \frac{\partial C_{\on}}{\partial \vy_{i_{*}}}=\frac{\partial C_{1}}{\partial \vy_{i_{*}}} + \frac{\partial C_{2}}{\partial \vy_{i_{*}}} + \frac{\partial C_{3}}{\partial \vy_{i_{*}}}.
\end{align}
However, since the embeddings of old data remain unchanged when we receive new samples, the gradient computation of $\vy_{i_{*}}$ doesn't involve $\vy_{i}$ and $C_{1}$. Therefore, we have $\frac{\partial C_{1}}{\partial \vy_{i_{*}}}=0$.
Furthermore, the gradient of the cost function $C_{2}$ with respect to $\vy_{i_{*}}$ is given by:
\begin{align}\label{eq:c2}
\begin{split}
\frac{\partial C_{2}}{\partial \vy_{{i_{*}}}} &= 2\sum_{j} {p_{{i_{*}}j}\tau_{{i_{*}}j}^{L} (\vy_{{i_{*}}} - \vy_j)} \\
&-(2 \sum_{j} (\tau_{{{i}_{*}}j}^{L})^{2} (\vy_{{i_{*}}} - \vy_j) + 4\sum_{{j_{*}}} (\tau_{{i}_{*}j_{*}}^{L})^{2} (\vy_{{i_{*}}} - \vy_{{j_{*}}}) \Big)\sum_{{i_{*}},j} \frac{p_{{i_{*}}j}}{Z}.
\end{split}
\end{align}
where $Z={\sum\limits_{k\ne l}\tau^{L}_{kl}+\sum\tau^{L}_{kl_{*}}+\sum\limits_{k_{*}\ne l_{*}}\tau^{L}_{k_{*}l_{*}}}$ is a global normalization constant. 
The gradient of remaining $C_{3}$ can be computed by
\begin{align}\label{eq:c4}
\begin{split}
\frac{\partial C_{3}}{\partial \vy_{{i_{*}}}} &= 4\sum_{{j_{*}}} {p_{{i_{*}}j_{*}}\tau_{{i_{*}}j_{*}}^{L} (\vy_{{i_{*}}} - \vy_{{j_{*}}})} \\
&\quad - (2 \sum_{j} (\tau_{{i_{*}}j}^{L})^{2} (\vy_{{i_{*}}} - \vy_j) + 4 \sum_{{j_{*}}} (\tau_{{i_{*}}j_{*}}^{L})^2 (\vy_{{i_{*}}} - \vy_{{j_{*}}})\Big)\sum_{{i_{*}},{j_{*}}} \frac{p_{{i_{*}}j_{*}}}{Z}.
\end{split}
\end{align}

\subsubsection{Algorithm of online t-SNE}
In this section, we introduce how to perform the algorithm of online t-SNE  
by optimizing the online KL divergence. This algorithm leverages the embedding of old data to learn the embedding of new data, proving highly effective when a dataset is sequential. \fref{process} shows the learning process of online t-SNE for this algorithm. 
A framework diagram outlining online t-SNE is presented in \fref{framework}. The distinctive strength of this algorithm lies in its capacity to offer continuous and adaptive high-resolution cell-type visualization. \fref{framework} shows that online t-SNE ensures the sequential understanding of new scRNA-seq data and the computational alignment between the embeddings of old and new data. Note that for a single scRNA-seq dataset, online t-SNE reduces to offline t-SNE and they have the same learning effect. The online t-SNE is general, data agnostic, and compatible with other advances of t-SNE. We present the learning procedure of online t-SNE in \aref{online-tSNE}.

The learning of online t-SNE is to iteratively minimize the online KL divergences via gradient descent. 
Online t-SNE achieves this by computing coordinates of low-dimensional embeddings for high-dimensional scRNA-seq samples, strategically positioning similar and dissimilar points closely and at a distance, respectively, within the dimensionally reduced map. This learning results in an effective visualizable embedding of the original high-dimensional scRNA-seq samples while preserving the relationships between them. 
The resultant joint embeddings of both old and new samples not only effectively reveal the clustering structure of old samples but also indicate the evolving behavior of new samples. For online t-SNE, discovering diversified evolving cluster structures is accomplished without the need for learning in two distinct manners. 

\begin{algorithm}[!h]
\caption{Online t-SNE}
\label{algo:online-tSNE}
\renewcommand{\algorithmicrequire}{\textbf{Input:}}
\renewcommand{\algorithmicensure}{\textbf{Output:}}
\begin{algorithmic}[1]
    \Require  old high dimensional samples $X=[\vx_{1}, \vx_{2},...,\vx_{n}]^{\top}$, new high dimensional samples $X_{*}=[\vx_{1_{*}},\vx_{2_{*}},...,\vx_{m_{*}}]^{\top}$, low-dimensional embeddings of old data $Y=[\vy_{1},\vy_{2},..,\vy_{n}]^{\top}$, perplexity $\mathrm{Perp}$, number of iterations $T$, leaning rate $\eta$, and momentum $\alpha(t)$. %%input
    \Ensure $Y_{*}^{(T)}=[\vy_{1_{*}}, \vy_{2_{*}}, ..., \vy_{m_{*}}]^{\top}$.    %%output
    \State  Compute pairwise affinities $p_{j|i_{*}}$ and $p_{j_{*}|i_{*}}$ with perplexity $\mathrm{Perp}$;
    \State Set $p_{ij_{*}}$ and $p_{i_{*}j_{*}}$ using \eref{HGD-old} and \eref{HGD-new}, respectively;
    \State Initialize $Y_{*}^{(0)}=[\vy_{1_{*}}, \vy_{2_{*}}, ..., \vy_{m_{*}}]^{\top}$ using $k$ nearest neighbours;
    \For{$t=1$ to $T$}
    \State	Compute low-dimensional affinities $q_{i_{*}j}$ and $q_{i_{*}j_{*}}$;
    \State	Compute gradient $\frac{\partial C_{\on}}{\partial \vy_{*}}$;
    \State	Set $\vy_{*}^{(t)}=\vy_{*}^{(t-1)}+\eta \frac{\partial C_{\on}}{\partial \vy_{*}}+\alpha (t)(\vy_{*}^{(t-1)}-\vy_{*}^{(t-2)})$;
    \EndFor
\end{algorithmic}
\end{algorithm}

On the other hand, to enhance the computational efficiency of joint probabilities, we simplify the computation steps by utilizing matrix operations. 
We employ automatic differentiation to obtain the chain rule of partial derivatives of $\vy_{*}$. 
Automatic differentiation refers to the automatic computation of the gradient vector of a cost function concerning each of its parameters. This functionality enables the automatic computation of gradients for low dimensional embeddings of new data in online t-SNE.
Specifically, we implement the online t-SNE algorithm by using Pytorch, because of its scalability and automatic differentiation routines. We consider Adam \cite{kingma2014adam} as the optimizer of online t-SNE 
to enhance the accuracy and efficiency of the algorithm.  
Adam can adjust the learning rate of $\vy_{i_{*}}$ based on the first-order moment estimate (mean) and second-order moment estimate (variance) of the gradient.   
This allows for a better adaptation to the characteristics of the data.

\subsubsection{Computation complexity of online t-SNE} 
Existing variants of offline t-SNE, usually exhibit computational and memory complexities that grow quadratically with the number of high dimensional samples. The evaluation of similarities in offline t-SNE incurs a time complexity of $\mathcal{O}(n^{2})$. This renders the application of the offline t-SNE impractical for datasets with more than $10000$ samples. 
To address this limitation, several approximations can be employed to significantly reduce the time complexity of offline t-SNE. 
The Barnes-Hut approximation efficiently reduces the computation complexity of offline t-SNE from $\mathcal{O}(n^{2})$ to $\mathcal{O}(n\log n)$. However, offline t-SNE with the Barnes-Hut approximation performs well for sample sizes up to $100000$  but slows down for larger datasets. An alternative technique interpolates repulsive forces on an equispaced grid and utilizes the fast Fourier transform (FFT) to accelerate the interpolation \cite{linderman2019fast}. This reduces the computation complexity of offline t-SNE to $\mathcal{O}(n)$ for sample sizes more than millions. 
Therefore, given $n$ historical samples and $m$ new samples, retraining offline t-SNE with exact inference, Barnes-Hut approximation, and FFT interpolation on the set of all sequential data need computation complexities with $\mathcal{O}((n+m)^{2})$, $\mathcal{O}((n+m)\log(n+m))$, and $\mathcal{O}(n+m)$, respectively. 

However, for online t-SNE, we do not need retraining t-SNE on the set of all sequential data. Online t-SNE only focuses on optimizing the embeddings of new data with size $m$. Therefore, the computation complexity of online t-SNE without approximation is $\mathcal{O}(m^{2})$ and mainly depends on $m$ other than $n$. However, for sequential data, we usually have $m$ much smaller than $n$. Online t-SNE has very good compatibility with offline t-SNE in terms of initialization and optimization methods. 
In particular, all existing approximation methods 
developed for offline t-SNE can be employed for online t-SNE. In other words, we can further reduce the computation complexity of online t-SNE to $\mathcal{O}(m)$.

\section{Experiments}
\subsection{Overview of online t-SNE}

Online t-SNE is an innovative single-cell embedding technique that accommodates both offline- and online sequential scRNA-seq analyses. It capitalizes on advanced stochastic neighbor embedding theory to project old data and new data into a unified low-dimensional embedding space. 
The online t-SNE endeavors to distill insights from old data to enhance its visualization performance in subsequent new data. 
Online t-SNE unfolds as a sequence of interconnected scRNA-seq subsets, with each subset involving a dynamic experiment. We show the framework and learning process of online t-SNE in \fref{framework} and \fref{process}, respectively.
Within each online subset, the algorithm commences by receiving a subset of samples, subsequently visualizing the patterns of its embedding. 
As the subset concludes, the algorithm acquires the correct embedding of scRNA-seq samples, harnessing this newfound knowledge to enhance and refine subsequent visualization of new samples. Online t-SNE merges the embeddings of old and new scRNA-seq data into a unified continuum. Diverging from existing t-SNEs that predominantly emphasize offline visualization of old data, online t-SNE treats both old data and future data as observations of embedding space, enabling the visualization of sequential scRNA-seq through a unified process. 
This iterative, on-the-fly learning process distinguishes online t-SNE as an agile and dynamic approach for sequential scRNA-seq, where adaptability and real-time understanding take precedence. 

Online t-SNE includes a range of pivotal steps, encompassing constructing compositional high-dimensional joint probability, representing the compositional low-dimensional joint probability of embedding space, composing online KL divergences, and minimizing the dissimilarity between the high-dimensional and low-dimensional joint probabilities.
In online t-SNE, visualization of sequential scRNA-seq samples involves the alignment between the compositional high-dimensional and low-dimensional probabilities. 
Online KL divergence serves a triple purpose: it effectively encapsulates the distribution of old scRNA-seq samples, connects the distributions of old and new scRNA-seq samples, and keeps the cellular heterogeneity of new scRNA-seq samples. 
Importantly, the joint embeddings are achieved without the reliance on retraining of full data including old data and new data, as it unifies knowledge of old data and new data. 
Crucially, the process of embedding learning exhibits inherent flexibility, rendering online t-SNE adaptable to a diverse array of scRNA-seq tasks.
In the following experiment, we use the term \textit{offline} t-SNE to denote existing variants of t-SNE because they can only \textit{offline} learn embedding of static scRNA-seq data. 

\begin{figure}[hb!]
    \centering
    \renewcommand{\tabcolsep}{1.0mm}    
    \begin{subfigure}[t]{1.0\linewidth}

    \centering\figclip{5 21 10 65}{0.8}{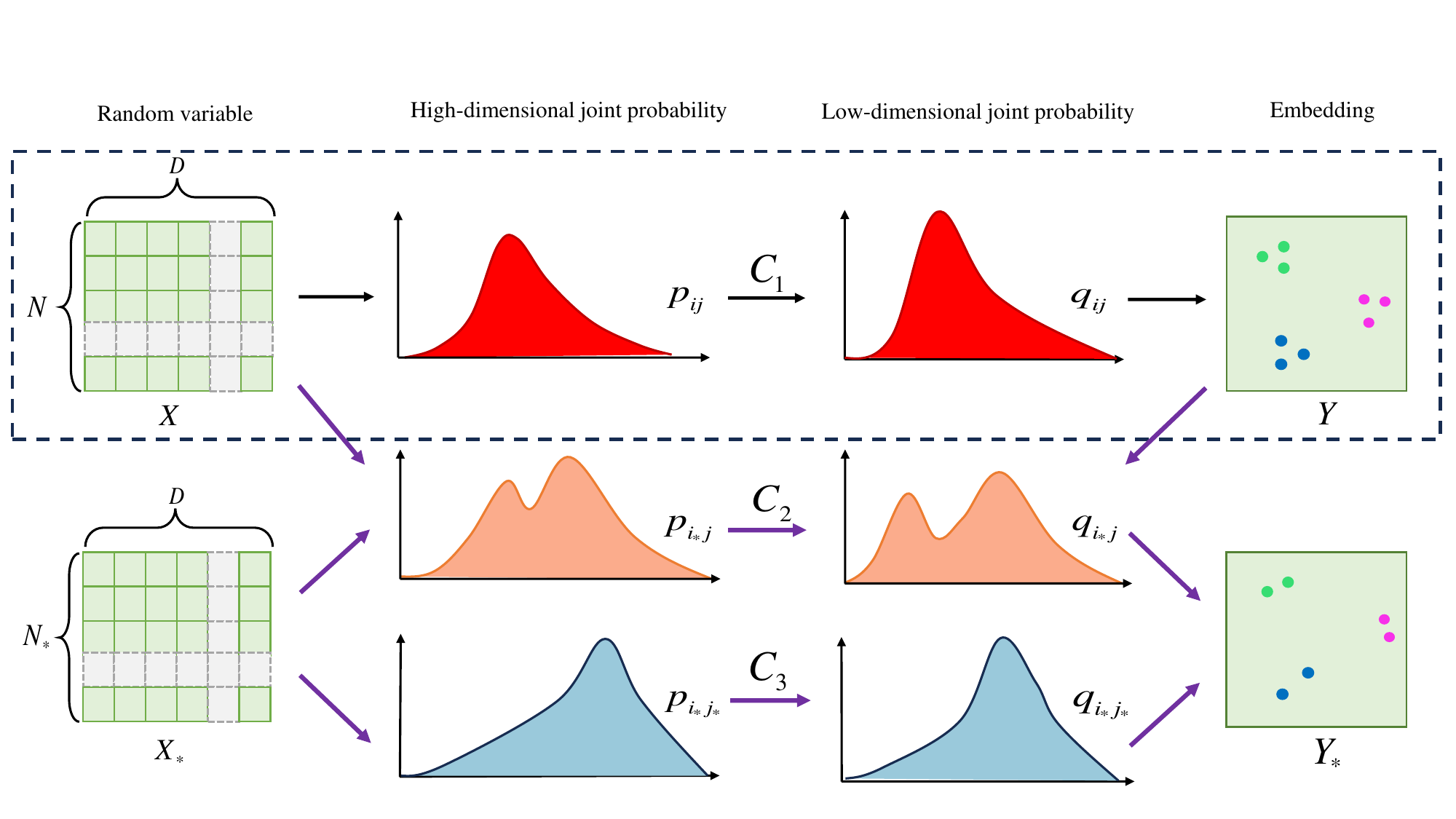} 
    \end{subfigure}
    \caption{Framework of online t-SNE. 
    For online t-SNE, there are three joint probabilities on the high-dimensional data spaces, $p_{ij}$ (in red) of old data $X\in\mathbb{R}^{D}$, $p_{i_{*}j}$ (in orange) between old data and new data $X_{*}\in\mathbb{R}^{D}$, and $p_{i_{*}j_{*}}$ (in blue) of new data. Correspondingly, there are three joint probabilities on the low-dimensional embedding spaces, $q_{ij}$, $q_{i_{*}j}$, and $q_{i_{*}j_{*}}$, which approximate $p_{ij}$, $p_{i_{*}j}$, and $p_{i_{*}j_{*}}$, respectively. $\{C_{i}\}_{i=1}^{3}$ denote the costs of online KL divergences. On the left, we have two subsets of scRNA-seq data (denoted by green matrices), including the old data subset $X$ and the new data subset $X_{*}$. 
    On the right, there are two low-dimensional embedding spaces learned by online t-SNE, $Y\in\mathbb{R}^{2}$ and $Y_{*}\in\mathbb{R}^{2}$, for $X$ and $X_{*}$, respectively. The black and purple arrows represent the knowledge flow from high-dimensional old data and new data to their low-dimensional embeddings, respectively. 
    $\{C_{i}\}_{i=1}^{3}$ denote the costs between the probabilities of high-dimensional data and the probabilities of low-dimensional embedding. Green, blue, and magenta points in $Y$ and $Y_{*}$ denote the low-dimensional embeddings of scRNA-seq samples.  Offline t-SNE (within a dashed rectangle) is a special case of online t-SNE, solely focusing on $p_{ij}$ and $q_{ij}$.
    }
    \label{fig:framework}
\end{figure}

\begin{figure}[hb!]
    \centering
    \renewcommand{\tabcolsep}{1.0mm}    
    \begin{subfigure}[t]{1.0\linewidth}
    \centering\figclip{150 83 300 60}{0.8}{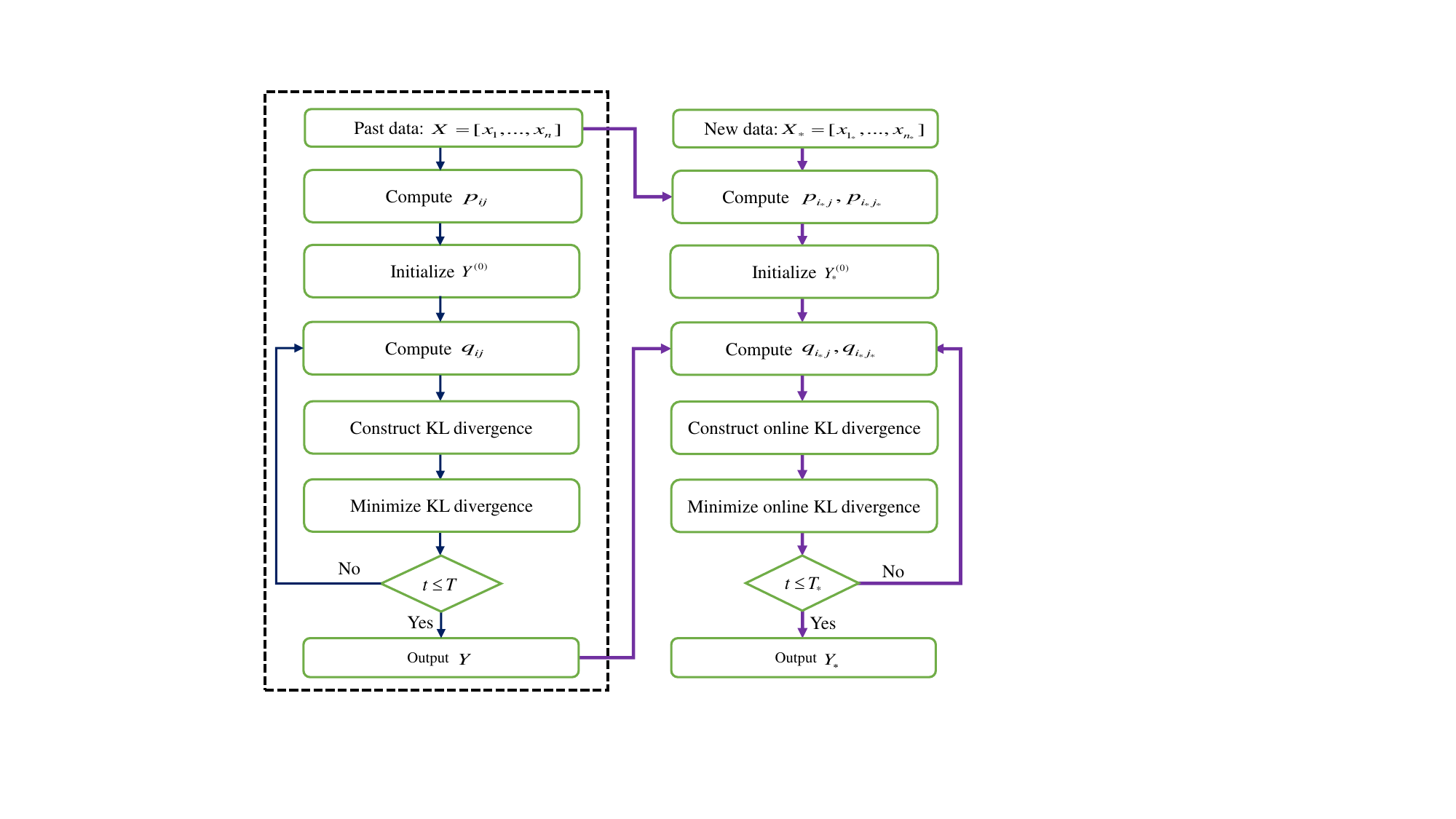}
    \end{subfigure}
    \caption{Learning process of online t-SNE. Given the low-dimensional embeddings $Y$ of old data, online t-SNE iteratively visualizes $X_{*}$ in real-time and avoids time-consuming retraining on the set $\{X, X_{*}\}$.
    Online t-SNE iteratively handles new data $X_{*}$ in real-time, leveraging the low-dimensional embeddings $Y$ of old data and avoiding time-consuming retraining on the set $\{X, X_{*}\}$. Offline t-SNE (within a dashed rectangle) learns the low-dimensional embeddings $Y$ of old data $X$. On the right, online t-SNE incorporates old data and its embeddings to learn new low-dimensional embeddings $Y_{*}$ of new data $X_{*}$. The blue and purple arrows denote the data processing direction of offline t-SNE and online t-SNE, respectively. In particular, the purple arrow (middle and top) signifies the data interaction between old data and new data to compute the high-dimensional joint distribution $p_{i_{*} j}$. Another purple arrow (middle and bottom) signifies the embedding interaction between the embeddings of old data and new data to compute the low-dimensional joint distribution $q_{i_{*} j}$. 
    }\label{fig:process}
\end{figure}
 
\subsection{Offline t-SNE struggles to visualize synthetic sequential dataset}

To highlight the inherent limitations of the offline t-SNE in discovering the clustering structures of sequential data, we begin with a simple illustrative online visualizing task with a synthetic sequential dataset. This synthetic sequential dataset comprises samples organized into two sequential subsets. We deliberately split the dataset so that samples from two subsets are separated. 
This task embodies an online visualization and analysis demand akin to what's often encountered in sequential scRNA-seq data.
In this synthetic experiment, sequential samples are randomly drawn from a $10$ dimensional Gaussian distribution (GD). All samples are divided into $10$ distinct and non-overlapping clusters by controlling the mean and covariance of the GD during sampling. 
There are $n=1000$ synthetic samples in total. Specifically, 
the samples of cluster $i$ are drawn from a GD with a covariance matrix of $\Sigma_{i}=I_{10}$ and a mean of $\mu_{i}$.
To simulate the sequential generation of samples, we randomly selected $700$ samples as the old data subset and the remaining $300$ samples as the new data subset. Notably, the subsets of old and new data follow the same distribution. 

However, when learning sequential data, existing t-SNE has to retrain on all collected data including old and new data. 
The number of retraining times depends on the generation frequency of new data. This could introduce high computational costs when sequential data arrives continuously and grows with time.  An alternative solution for the visualization of sequential data is to independently perform t-SNE on the subsets of old data and new data, respectively. Therefore, we can obtain two subsets of embeddings for the old and new data, respectively. Furthermore, we may be suggested to merge the separate embeddings of the old and new data as joint embeddings to visualize both old and new data together. However, this method leads to discrepancies between samples from the same class but located in different subsets.

\begin{figure}[h!]
    \centering
    \renewcommand{\tabcolsep}{1.0mm}
    \scriptsize
    \begin{tabular}{p{0.5mm}*{3}{c}}
        & \figclip{88 10 83 10}{0.32}{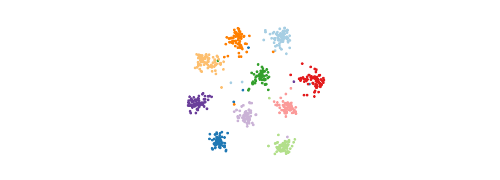}
        & \figclip{83 15 80 10}{0.30}{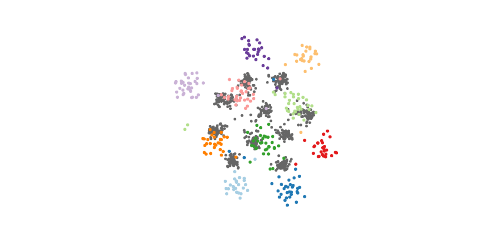} 
        & \figclip{75 15 65 15}{0.33}{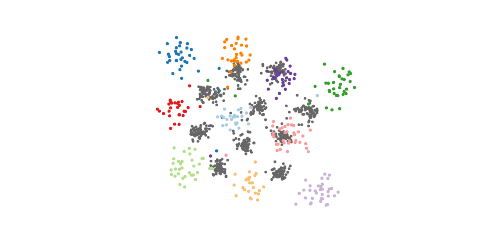} \\
        & (a) & (b) & (c) 
    \end{tabular}
    \caption{Visualization of synthetic sequential data: (a) visualization of the old data subset ($n$=700) using offline t-SNE, (b) visualization of the new data subset ($n$=300) using offline t-SNE with random initialization, (c) visualization of the new data subset using offline t-SNE with PCA initialization. The shade points shown in subplots (b) and (c) denote embeddings of the old data. We use 10 colors to label 10 different clusters, respectively. offline t-SNE cannot align the embeddings of the new data in both subplots (b) and (c) with the embeddings of the old data in subplot (a).}
    \label{fig:toy}
\end{figure}

As shown in \fref{toy}, we illustrate that offline t-SNE fails to embed and visualize synthetic sequential data in an online scenario.
We perform two separate offline t-SNEs successively learning on two subsets: the old and new data. 
Two separate offline t-SNEs may untangle the clustering structures associated with different subsets from those that are subset-invariant within data, but have difficulty in capturing the subset-invariant clustering structures shared between old data and new data.
In \fref{toy}, subplots (b) and (c) show that the cluster structures independently learned by separate offline t-SNEs on new data significantly differ from the cluster structures of old data in subplot (a). 
The embedding structures in subplots (b) and (c) exhibit a lot of discrepancies in terms of the positions of clusters, the relative position between different clusters, and the relative position between the clusters of new data and old data. 
Offline t-SNE cannot connect the embedding of old data and new data without retraining on the full data from scratch. Offline t-SNE fails to yield a convincable and consistent cluster structure of new data. For instance, the purple cluster of old data in subplot (a) is located on the left side of the 2D embedding space, while the purple cluster of new data in subplot (a) and subplot (c) is located on the top side. This means the position of the purple cluster in the 2D embedding space has changed from old data to new data. This experiment confirms that offline t-SNE struggles to obtain comparable and consistent embeddings for different subsets.

\subsection{Learning consistent embedding of mouse neocortex cell dataset}
To assess online t-SNE's efficacy in learning the consistent embedding structure of real-world sequential scRNA-seq data, we applied online t-SNE to mouse neocortex cell dataset \cite{tasic2018shared}. 
This dataset is collected from the adult mouse neocortex and includes $23,822$ individual samples including $14,249$ cells from the primary visual cortex and $9,573$ cells from the anterior lateral motor (ALM) cortex. 
Understanding different types of brain cells and their roles in circuits is a key step in figuring out how the brain manages different tasks. 
The samples are categorized into $133$ transcriptomic types. In this experiment, we meticulously follow the procedures in \cite{kobak2019art} to preprocess the dataset. 
To clearly compare the visualization difference and embedding performance between offline t-SNE and online t-SNE, we selected transcriptomic types with distinguishable clusters obtained by standard t-SNE, which means there is a high degree of distinction between these types. By employing these highly distinguishable types, we can avoid the confusion caused by the overlapped cluster ranges. 
Specifically, there are $14$ transcriptomic types and $6000$ samples used in this experiment. 
By exploring the correspondence between transcriptomic types and embeddings, we can analyze the clustering behavior of cells in terms of activity patterns and information processing. 

In \fref{art}, we present 
the embedding (subplot (a)) of the original full data using offline t-SNE and the embedding (subplot (b)) of the old and new data using online t-SNE.  
The embedding of the original full data is obtained by training an offline t-SNE on all data. 
In subplot (a), all pairs of clusters captured by offline t-SNE are separable.  The embedding of the old and new data is acquired by performing online t-SNE on separate old and new data.
In subplot (b), cell samples from the new data are plotted with the same color with the embedding of offline t-SNE. Online t-SNE excels in detecting clusters with a given few new samples, effectively segregates biological and technical variation, and provides clearer cluster boundaries. 
We observe two-level consistencies between the embeddings of old and new data: global and local consistencies. The first is the consistency of global structures between old data and new data. All relative distances between clusters in the embedding of old data are successfully transferred in the embedding of new data. The second is the consistency of single local clusters between old data and new data. 
Significantly, the overall embedding of new data using online t-SNE is perfectly consistent with the embeddings of the old data and collected full data. 
These consistencies indicate that online t-SNE has a comparable learning capacity with offline t-SNE on full data. However, the online t-SNE does not need retraining and has higher learning efficiency. This experiment provides robust evidence to verify the effectiveness of online t-SNE on real-world mouse cortex data.

\begin{figure}[h!]
    \centering
    \renewcommand{\tabcolsep}{2.0mm}
    \scriptsize
    \begin{tabular}{p{0.5mm}*{2}{c}}
        & \figclip{25 5 25 2}{0.40}{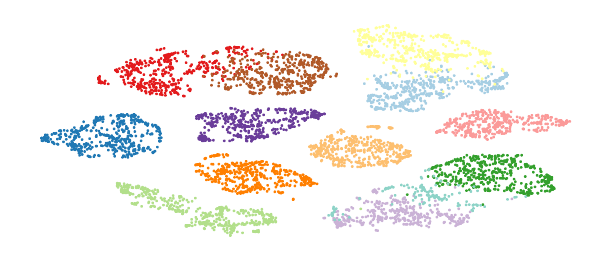}
        & \figclip{25 5 25 2}{0.40}{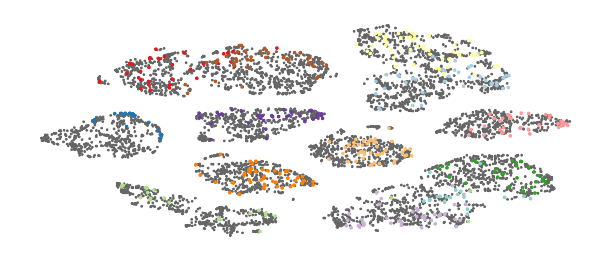} \\
        & (a) & (b) 
    \end{tabular}
    \caption{Visualization of adult mouse neocortex cell dataset: (a) visualization using offline t-SNE on the collection of old and new data. (b) visualization using online t-SNE on the new data.  The shade nodes shown in subplot (b) denote the learned embeddings using offline t-SNE on old data. Cell types are colored by: Pvalb Gpr149 Islr, light blue; CR Lhx5, dark blue; L2/3 IT ALM Macc1 Lrg1, light green; L5 IT ALM Lypd1 Gpr88, dark green; L5 IT VISp Whrn Tox2, pink; Lamp5 Plch2 Dock5, red; L6 IT ALM Oprk1, orange; L2/3 IT VISp Adamts2, orange-yellow; L5 IT ALM Npw, light purple; Sst Myh8 Fibin, dark purple; Pvalb Reln Itm2a, light yellow; Lamp5 Ntn1 Npy2r, brown; and L5 IT ALM Pld5, turquoise. Online t-SNE on the new data can leverage the learned embeddings of old data without retraining from scratch.}
    \label{fig:art}
\end{figure}

\subsection{Mitigating the batch effect of kidney cell dataset}

\begin{figure}[h!]
    \centering
    \renewcommand{\tabcolsep}{2.0mm}
    \scriptsize
    \begin{tabular}{p{0.5mm}*{2}{c}}
        & \figclip{0 0 5 2}{0.45}{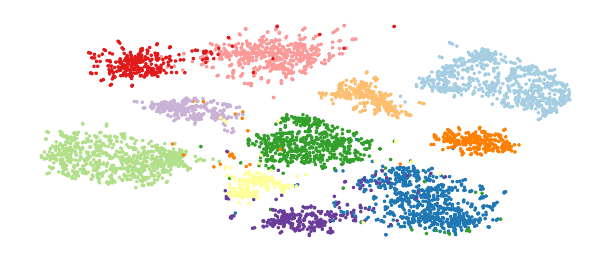}
        & \figclip{15 0 10 2}{0.45}{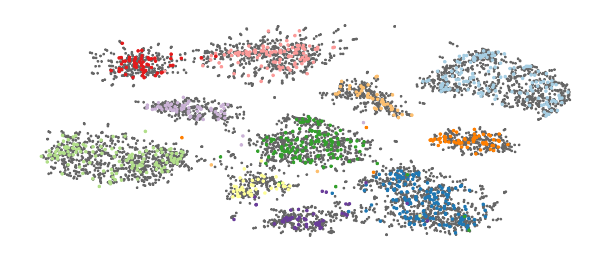} \\
        & (a) & (b) 
    \end{tabular}
    \caption{Visualization using online t-SNE on renal cell dataset with two batches: (a) visualization of the first batch. (b)  visualization of the second batch. Colors denote the types as: PTC, light blue; LoH.TAL, dark blue; CD4.T.cells, light green; IC.A, dark green; Macro., pink; DC, red;  LoH.DTL, orange; vSMC, orange-yellow; EC.glom, light purple; LoH.ATL, dark purple; and PC.CD, light yellow.}
    \label{fig:renal}
\end{figure}

In single-cell studies, it is often necessary to successfully conduct multiple independent experiments and therefore gather all experimental data 
to obtain a sufficient size of samples for effective visualization and analysis. Unfortunately, batches of scRNA-seq data may suffer from batch effects \cite{hicks2018missing}, which are especially challenging in the extensive studies of human tissues. In this experiment, we aim to alleviate the batch effect of kidney cell dataset \cite{quatredeniers2023meta} by using online t-SNE. Kidney cell data inherently carry batch effects due to various sample generations, leading to potentially different representations of embedding space. 
Notably, scRNA-seq studies of kidney cells employ various technologies, miscellaneous data preprocessing, and experimental workflows.  This lack of consensus diminishes the transferability of knowledge and reliability of understanding between batches.
Effective mitigation of batch effects is crucial for facilitating meaningful understanding and comparisons between kidney cell samples, achieved through powerful unsupervised clustering and visualization technology. 
We present an online analysis of kidney cell batches via online t-SNE. Cell types were assigned to clusters using broad cell type markers, consensus cell type signatures were computed, and the labeled cell was integrated to map cell types through the statistical relationship between one batch and another batch. 

In this scenario, batch effects of kidney cells may introduce varied differences in the data, such as different global and local structures of embedding. There are two batches of kidney cell samples. The first batch has $5000$ samples and the second batch has $1000$ samples. The two batches have the same cell types. 
We explored two clustering approaches for comparison: offline t-SNE and online t-SNE. Subsequently, PCA was computed, and the first $2$ principal components (PCs) were employed for the initialization of low-dimensional embedding. 
As shown in \fref{renal}, online t-SNE effectively mitigates the batch effect and successfully aligns the clusters of one batch with the clusters of another batch. 
In this experiment, we have  $11$ transcriptomic types of kidney cell samples.
Despite the uneven distribution of clusters, the cell type of cluster exhibited high consistency across the two batches of kidney cells. Compared to subplot (a) and subplot (b), their cluster distances, ranges, and positions are highly similar due to the knowledge transferring via $p_{i_{*}, j}$ and $q_{i_{*}, j}$. 
Multiple CD4.T cells in the second dataset (subplot (b)) were effectively projected into the embedding space (subplot (a)) of the first dataset. 
Ultimately, we demonstrate that online t-SNE facilitates cell-type clustering and understanding across batches of kidney cells.
It is noteworthy that online t-SNE stands out as a highly effective method for correcting batch effects.

\subsection{Exploring shared embedding across diversified pancreatic cell dataset}
In this experiment, we explore evolving and diversified pancreatic cell data, perform high-resolution unsupervised clustering, allocate cell types to clusters based on known markers, and compute consensus cell type signatures. scRNA-seq has gained significant traction in pancreatic research. 
While major pancreatic cell types are generally discerned, the issue of inconsistent cell type identification persists across diversified data sources, leading to a lack of reproducibility.
It is reasonable to expect variations in transcriptomic signatures across different diversified pancreatic cell data sources. The limited sample size in most pancreatic single-cell data sources, influenced by the technology's cost and the scarcity of healthy human samples, hampers the generalizability of conclusions to the broader population. 
To improve the generalizability of findings to broader human populations, it is often encouraged to explore various datasets that include diversified samples from different experiments. Here, we select the pancreatic cell datasets including the diversified Baron dataset \cite{baron2016single} and Xin dataset \cite{xin2016rna} to demonstrate the learning capacity of online t-SNE. In this experiment, Baron data and Xin data are heterogeneous.

To visualize diversified pancreatic samples in shared low dimensional-embedding space, we compare the cell type clusters learned by offline t-SNE on the collection of diversified pancreatic datasets with the clusters learned by online t-SNE. 
As shown in \fref{pancreas}, subplot (a) denotes the clusters learned by offline t-SNE containing points with two colors, various sizes, and shapes. 
The two pancreatic cell datasets (identified by two colors) in the embedding space have no intersection, which means no shared embedding is discovered by offline t-SNE. 
Subplot (b) denotes the clusters learned by online t-SNE on the Baron dataset.
Subplot (c) denotes the clusters learned by online t-SNE on the Xin dataset. 
The clusters in subplot (c) are aligned with the clusters in subplot (b) in terms of cell type and cluster range. 
This alignment between subplot (b) and subplot (c) demonstrates that online t-SNE successfully captures the shared embeddings between two pancreatic cell datasets. This experiment confirms the eagerness of visualization on diversified pancreatic cell data through shared embedding.
The advantages of online t-SNE over offline t-SNE include reduced dissociation bias and the ability to absorb knowledge of other data sources for understanding shared patterns of diversified samples.  
Ultimately, online t-SNE facilitates diversified cell type clustering and visualization, enhancing reproducibility and reliability in the future scRNA-seq analysis of pancreatic cells.

\begin{figure}[h!]
    \centering
    \renewcommand{\tabcolsep}{1.0mm}
    \scriptsize    
    \begin{tabular}{p{0.5mm}*{3}{c}}
        & \figclip{20 0 20 2}{0.32}{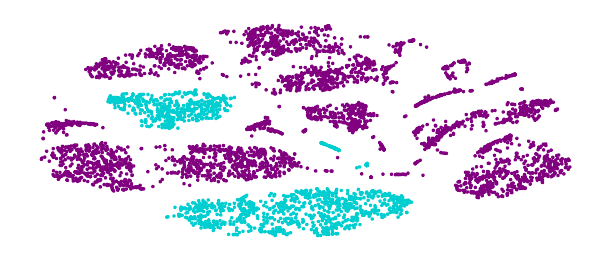} 
        & \figclip{20 0 20 2}{0.32}{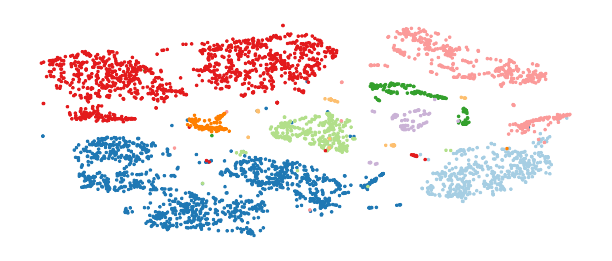}
        & \figclip{20 0 20 2}{0.32}{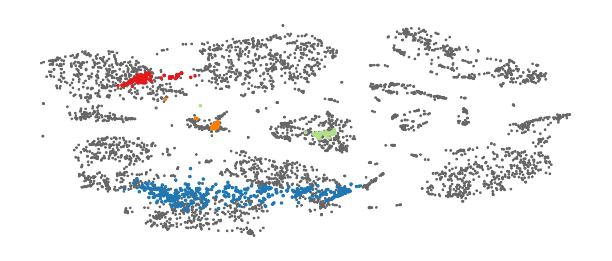} \\
        & (a) & (b) & (c)
    \end{tabular}    
    \caption{Visualization on the shared embedding of the diversified pancreatic cell datasets: (a) visualization using offline t-SNE on the collection of Baron and Xin pancreatic cell dataset, (b) visualization using online t-SNE on Baron dataset, (c) visualization using online t-SNE on Xin dataset. In subplot (a), the dark purple points and cyan points denote the embeddings of the Baron dataset and Xin dataset, respectively. In subplot (b), the colorful points denote the embeddings of the Baron dataset. In subplot (c), the colorful points denote the embeddings of the Xin dataset. }
    \label{fig:pancreas}
\end{figure}

\subsection{Discovering new biological structures from mouse visual cortex cell dataset}
In general, sequential scRNA-seq data is subject to continuous updates. Just performing visualization on old data usually obtains limited information and understanding of new structures. For sequential scRNA-seq data, there are both heterogeneous and homogeneous structures. Homogeneous structures are shared between old and new data. The new structure can be seen as a heterogeneous structure that is unseen in old data and only included in new data. 
This problem is frequently observed in various sequential scRNA-seq tasks, where new data always contains additional cell types not present in the old data. These additional cell types are important and informative for the understanding of sequential scRNA-seq data. 
The challenge of discovering new structures of sequential scRNA-seq data timely is widely existing in the bioinformatics community. 
Visualization and analysis of new data with new structures encounter various difficulties, such as concept drift and representation sequential dependency between existing old structures and new structures. Stringent experimental practices and a well-designed experiment can minimize these difficulties. Misapprehending new structures is notorious for interfering with downstream analyses of sequential scRNA-seq data.
Furthermore, handling new structures without due care can lead to the loss of genuine biological signals in the new data. There is no such method has been developed to discover and visualize new structures in sequential scRNA-seq data. 

\begin{figure}[h!]
    \centering    
    \scriptsize
    \begin{subfigure}[t]{1.0\linewidth}
    \renewcommand{\tabcolsep}{0.5mm}
    \begin{tabular}{p{0.5mm}*{3}{c}}
        & \figclip{20 0 20 2}{0.33}{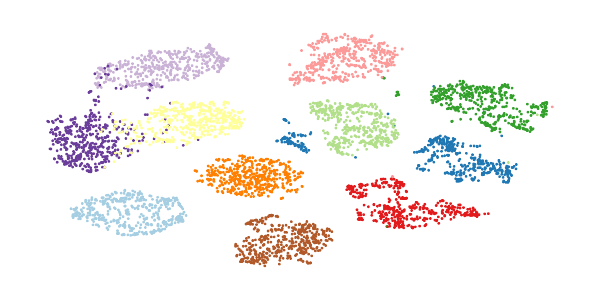} 
        & \figclip{20 0 20 2}{0.33}{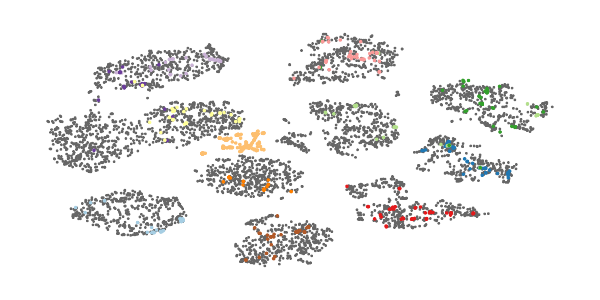} 
        & \figclip{10 0 12 2}{0.33}{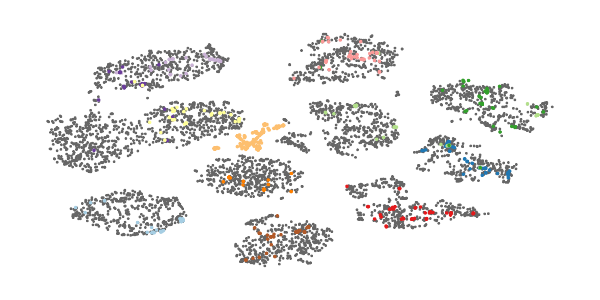} \\ 
        & (a) & (b) & (c) 
    \end{tabular}
    \end{subfigure}

    \caption{Visualization of new biological structures only appeared in new mouse visual cortex cells dataset (\cite{hrvatin2018single}): (a) the visualized embedding of old mouse cortex cell data, (b) the new biological structures (denoted by dark yellow) discovered by online t-SNE on new mouse cortex cell data, and (c) the visualization of the new biological structure through online t-SNE with a larger local exaggeration ($30$).
    Subplot (a) shows a dense distribution of $11$ colors, with each color representing a cell type of the old data. Subplot (b) has a denser distribution of $12$ colors, with an additional color denoting a new cell type that appeared in the new data. With the increase of local exaggeration, the embeddings of the new cell type change to a more compact cluster (shown in subplot (c)) in terms of cluster position, shape, and scope. }\label{fig:hrvatin}    
\end{figure}

\begin{figure}[h!]
    \centering    
    \scriptsize

    \begin{subfigure}[t]{1.0\linewidth}
    \renewcommand{\tabcolsep}{0.2mm}
    \begin{tabular}{p{0.5mm}*{3}{c}}
        & \figclip{10 0 10 2}{0.33}{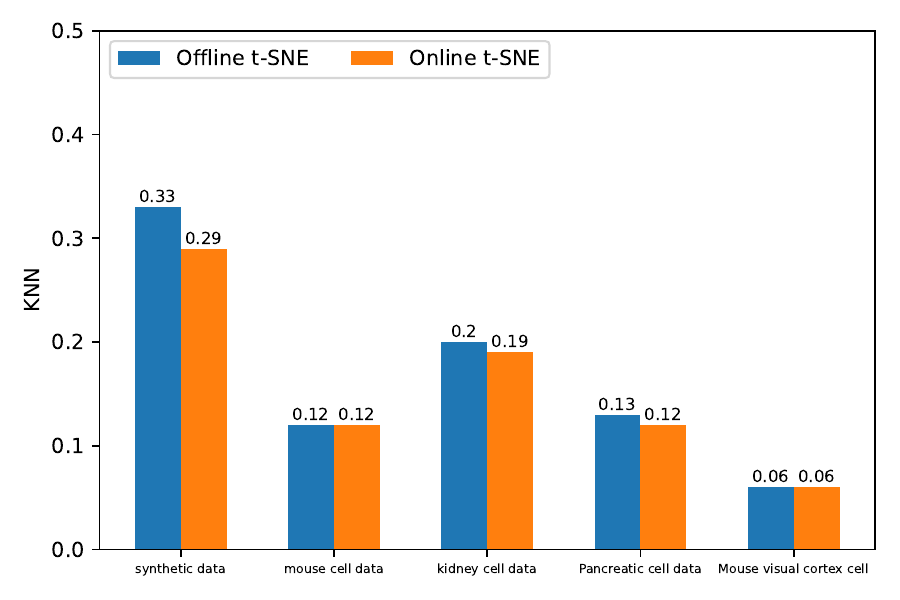} 
        & \figclip{10 0 10 2}{0.33}{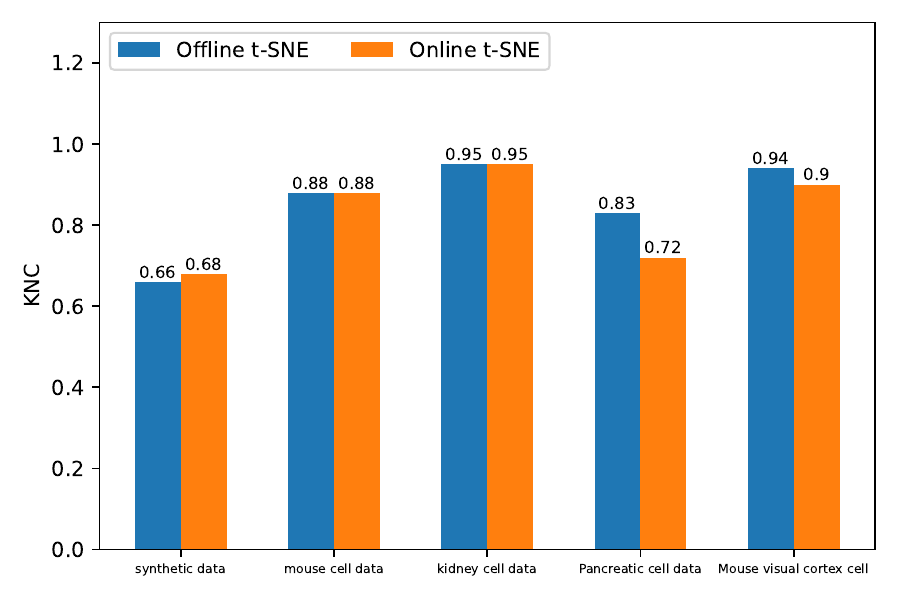}      
        & \figclip{10 0 10 2}{0.33}{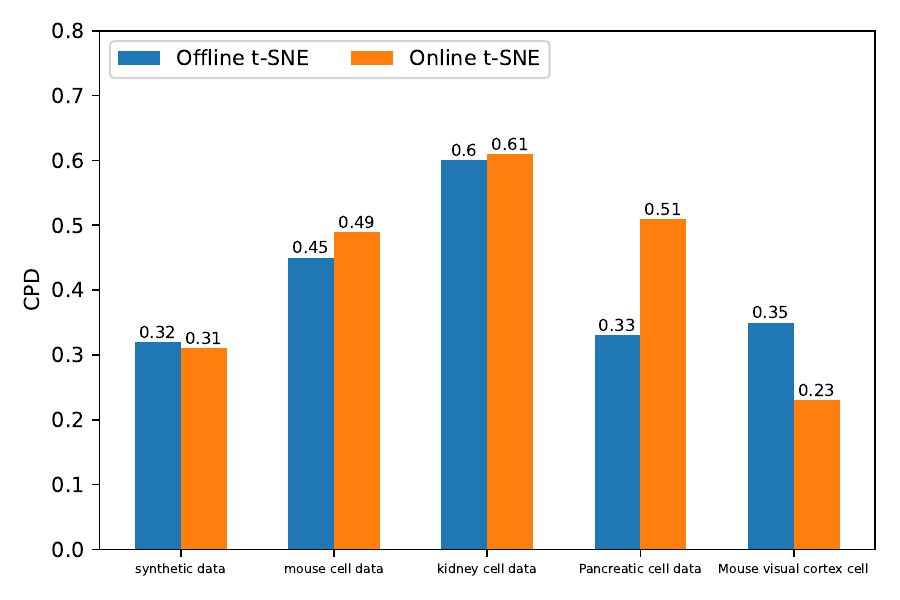} \\ 
        & (a) $\mathrm{KNN}$ & (b) $\mathrm{KNC}$ & (c) $\mathrm{CPD}$
        % & (a) & (e) & (f)
    \end{tabular} 

    \end{subfigure}
    \caption{Visualization performances of online t-SNE and offline t-SNE across various datasets: (a) $\mathrm{KNN}$ of online t-SNE and offline t-SNE, (b) $\mathrm{KNC}$ of online t-SNE and offline t-SNE, and (c) $\mathrm{CPD}$ of online t-SNE and offline t-SNE.}\label{fig:quality}    
 
\end{figure}

In this experiment, online t-SNE can unveil more cell types, particularly when they have not appeared in old data, thereby contributing to the identification of new cell types. 
To assess the capability of discovering new structures for online t-SNE, we employ the mouse visual cortex cells dataset curated by \cite{hrvatin2018single}. 
We separate the dataset into two sequential pieces: a piece of old data and another piece of new data. Note that the new data only contains one new cell type.
At first, we perform offline t-SNE on the old data to obtain its embeddings (shown in subplot (a) \fref{hrvatin}). Based on the obtained embeddings of old data, we continuously perform online t-SNE on the piece of new data and then acquire new embeddings. In \fref{hrvatin}, we show the visualization of old data as well as new data with new biological structures. 
In subplot (b) \fref{hrvatin}, we visualize the embeddings of new data. Using online t-SNE, the most of homogeneous structures of new data are accurately aligned with the embeddings of the old data. 
In addition to the known cell types (denoted by similar colors with subplot (a)) in the old data, subplot (b) shows that the new data also contains a new cell type (denoted by orange-yellow). Notably, the new cell type is successfully discovered by online t-SNE. In this case, the new cell type indicates a Micro$_2$ cell. 
In this context, a larger learning rate of online t-SNE is often required to discover new structures.
Furthermore, we introduce a local exaggeration technique for the similarities $q_{i_{*}, j}$ between new samples and old samples to strengthen the discovery of new structures.  
In particular, this local exaggeration technique does not impact the embeddings of old data.

\section{Discussion and Conclusion}
To assess the quality and reliability of the embeddings produced by online t-SNE and offline t-SNE, we employ three metrics used in \cite{kobak2019art}: $\mathrm{KNN}$, $\mathrm{KNC}$, and $\mathrm{CPD}$.  Specifically, $\mathrm{KNN}$ measures the proportion of $k$-nearest neighbors in the high-dimensional space that are successfully retained in the low-dimensional embedding space. It assesses the preservation of microscopic structures. 
$\mathrm{KNC}$ quantifies the proportion of successfully retained $k$-nearest neighbors for class means in both high- and low-dimensional spaces. This metric focuses on the preservation of relative positions after embedding concerning classes. It quantifies the retention of medium-scale structures.
$\mathrm{CPD}$ assesses global structure preservation by computing the Spearman correlation between pairwise distances in high- and low-dimensional embedding spaces. 
A high $\mathrm{CPD}$ indicates successful preservation of the global structure.
For all experiments, we compute the metrics (shown in \fref{quality}) for offline t-SNE and online t-SNE.  Note that the metrics of offline t-SNE are based on the embeddings of the full data including new data and old data. 
The metrics of online t-SNE and offline t-SNE are denoted as orange bars and blue bars, respectively. As shown in subplots (a), (b), and (c), the performances of online t-SNE are comparable with offline t-SNE in terms of $\mathrm{KNN}$, $\mathrm{KNC}$, and $\mathrm{CPD}$. Subplot (a) indicates that online t-SNE preserves microscopic structures well in the visualizations of the mouse neocortex cell dataset and the mouse visual cortex cell dataset. Compared with the offline t-SNE on full data, we only observe a little bit of $\mathrm{KNN}$ performance loss for online t-SNE on the synthetic dataset, kidney cell dataset, and pancreatic cell dataset. 
Subplot (b) shows that online t-SNE retains medium-scale structures well in the visualizations of the synthetic dataset,
the mouse neocortex cell dataset, and the kidney cell dataset. However, the $\mathrm{KNC}$ performances of online t-SNE on the pancreatic cell dataset and the mouse visual cortex cell dataset are slightly lower than offline t-SNE on the full data. 
Subplot (c) exhibits higher $\mathrm{CPD}$ of online t-SNE on the mouse neocortex cell dataset, the kidney cell dataset, and the pancreatic cell dataset than offline t-SNE, which reveals a better global structure preservation capacity of online t-SNE.
On the synthetic dataset and the mouse visual cortex cell dataset, the $\mathrm{CPD}$ performances of online t-SNE are more or less inferior to offline t-SNE. These quantitative metrics state that the quality of visualization using online t-SNE is competitive and reliable.

The rapid advancement of single-cell transcriptomics demands the development of corresponding computational online visualization and analysis algorithms to discover and extract integrative insights from extensive and diversified sequential scRNA-seq data. This discrepancy underscores the necessity for methods that overcome the limitations of popular offline t-SNE and can be seamlessly extended to a computational online analysis of sequential scRNA-seq data. Online t-SNE addresses this demand as a comprehensive and scalable approach for considering cellular heterogeneity and regulatory mechanisms. For online t-SNE, the distributions of old and new scRNA-seq samples are modeled as three joint probabilities, employing an online KL divergence to map old and new high-dimensional scRNA-seq samples into a shared low-dimensional embedding space. We demonstrate that online t-SNE not only captures the relationship between scRNA-seq samples of one subset but also the cross relationship between scRNA-seq samples of old and new subsets. 
Starting from the embeddings of the old data, online t-SNE updates the embeddings as well as the joint probability by incorporating new data. 
This not only utilizes the spatial information of the existing data but also adapts to changes in data distribution and structure. 
Thus, online t-SNE facilitates the simultaneous learning of heterogeneous scRNA-seq datasets and the discovery of new biological structures, advancing current offline t-SNE. Online t-SNE also avoids the usual reliance on integrated large datasets for structure understanding and discovery, mitigating the cost of data collection and the difficulty of data processing.

Our online t-SNE underwent extensive experiments across diverse visualization and analysis tasks of scRNA-seq data, achieving continual representation capacity that surpasses or rivals existing offline t-SNE. 
These experiments underscore the broad applicability and milestone of online t-SNE, eliminating the difficulty of sequential understanding and analysis of scRNA-seq data. 
The fact that offline t-SNE struggles to visualize sequential data is one of the well-known limitations \cite{wattenberg2016use}. Once the embedding learned by offline t-SNE is constructed, it is difficult to incorporate new data into the existing embedding without retraining the collection of all subsets and damaging the learned embedding. 
Online t-SNE ingeniously connects the embeddings of old data and new data, transfers the knowledge of old data to new data, and obtains continual visualization for future scRNA-seq data.  Online t-SNE is compatible with most t-SNE variants and can be seamlessly combined with them \cite{abdelmoula2016data,kobak2019art,linderman2019fast}. 
Moreover, it sets the stage for the evolution of t-SNE that does not solely rely on old and static data.  
Overall, simplicity and generality are major strengths of online t-SNE: it can be used as drop-in replacements for popular variants of offline t-SNE, with major benefits in efficiency and performance while retaining simple training and inference procedures. We have evidenced its broad applicability, and we anticipate that it will serve as a catalyst for the development of numerous visualization technologies and foster a more unified perspective on expressive t-SNE. 

However, despite its promising capabilities, online t-SNE is not without potential limitations and areas for improvement. Challenges may arise in noised sequential scRNA-seq data, such as samples with noise, as this requires additional control of similarity to accurately describe the three probabilities. The incorporation of multimodal could also enhance online t-SNE's capacity to analyze sequential multimodal datasets. 
We are optimistic that online t-SNE will alleviate the challenges associated with the development of emerging sequential single-cell datasets and tasks. 

\bibliographystyle{unsrt}  
\bibliography{online_tsne}

\end{document}